\documentclass{aastex}
\usepackage{emulateapj5}
\usepackage{apjfonts}
\usepackage{epsfig}

\slugcomment{ }

\begin{document}

\title{Early star formation traced by the highest redshift quasars}

\author{R. Maiolino\altaffilmark{1}, Y. Juarez\altaffilmark{2},
R. Mujica\altaffilmark{2}, N. M. Nagar\altaffilmark{3} and
E. Oliva\altaffilmark{1,4}}
\altaffiltext{1}{INAF -- Osservatorio Astrofisico di Arcetri, largo E. Fermi 5,
50125 Firenze, Italy; maiolino@arcetri.astro.it}
\altaffiltext{2}{Instituto Nacional de Astrof\'{i}sica \'Optica y Electronica,
 Puebla, Luis Enrique Erro 1, Tonantzintla, Puebla 72840,  Mexico;
yjuarez@inaoep.mx, rmujica@inaoep.mx}
\altaffiltext{3}{Kapteyn Institute, University of Groningen,
Landleven 12, 9747 AD Groningen, The Netherlands;
nagar@astro.rug.nl}
\altaffiltext{4}{Telescopio Nazionale Galileo, Calle Alvarez de Abreu, 70,
38700 Santa Cruz de La Palma, Spain;
oliva@tng.iac.es}

\begin{abstract}
The iron abundance relative to $\alpha$-elements
in the circumnuclear region of quasars is regarded as a
clock of the star formation history and, more specifically, of the
enrichment by type Ia supernovae (SNIa).
We investigate the iron abundance in a sample
of 22 quasars in the redshift range 3.0$<$z$<$6.4 by measuring their
rest frame UV FeII
bump, which is shifted into the near-IR, and by comparing it with the
MgII $\lambda$2798 flux. The observations were performed
with a device that can 
obtain near-IR spectra in the range 0.8--2.4$\mu$m in
one shot, thereby enabling an optimal removal of the continuum
underlying the FeII bump. We detect iron in all quasars including the
highest redshift (z=6.4) quasar currently known.
The uniform observational technique
and the wide redshift range allows a reliable study of the trend of the
FeII/MgII ratio
with redshift. We find the FeII/MgII ratio is nearly constant at all 
redshifts, although there is marginal evidence for a higher
FeII/MgII ratio in the quasars at z$\sim$6.
If the FeII/MgII ratio reflects the Fe/$\alpha$ abundance, this result
suggests that the z$\sim$6 quasars have already undergone a major episode
of iron enrichment. We discuss the possible implications of this finding
for the star formation history at z$>$6.
We also detect a population of weak iron
emitters at z$\sim$4.5, which are possibly hosted in systems that
evolved more slowly.
Alternatively, the trend of the FeII/MgII ratio at
high redshift may reflect significantly different physical conditions
of the circumnuclear gas in such high redshift quasars.
\end{abstract}


\section{Introduction}

The iron abundance relative to $\alpha$-elements has been
regarded as a clock indicator of the star formation history
\citep[e.g.][]{hamfer99}. Indeed
most of the iron in local galaxies is thought
to be produced by type Ia SNe, while $\alpha$-elements
are predominantly produced by type II SNe. The difference between the
progenitor masses of SNIa and SNII (hence in their life times)
translates into an enrichment delay between
the Fe and $\alpha$-elements, which was
generally thought to be about 1 Gyr \citep{wheeler89}.
However, more recent
studies have shown that SNIa are also enhanced in young stellar
systems \citep[e.g.][]{manet03}.
In particular the SNIa progenitors
may be as massive as 8~M$_{\odot}$ \citep{greren83},
which have a lifetime of $\sim 3\times 10^7$yr. As a consequence
the maximum of iron enrichment may occur on relatively short
timescales depending on the star formation history and on
the initial mass function \citep{matrec01}.

The prominent emission lines observed in quasars allow us to investigate
the metallicity of galactic nuclei at large cosmological distances
\citep{hamfer99}.
The main iron feature in quasar spectra is the UV bump
at 2200--3090\AA ,
which is the blend of several thousand FeII (and some FeIII) lines.
The ratio
of the FeII (UV) bump to the MgII $\lambda$2798 doublet is sensitive
to the Fe/$\alpha$ abundance ratio,
but is also sensitive to other
physical properties of the emitting region such as
microturbulence, density, and ionization parameter \citep{veret03}.
As a consequence it is not simple to quantitatively
derive the Fe/$\alpha$ ratio from the FeII(UV)/MgII ratio alone. The
FeII(UV)/MgII ratio can be used as a {\it relative} indicator of Fe/$\alpha$
among quasars, under the assumption that the physical
properties of the emitting regions are the same.

At z$>$3 the UV iron bump is shifted in the infrared. Several authors
have measured the FeII/MgII ratio in high redshift
quasars by means of near-IR spectroscopy \citep{aoki02,dieet02,dieet03,freet03,
iwaet02,thompson99}. Yet, the main difficulty in several
of these studies \citep[with the exception of ][]{dieet02,dieet03}
is that most infrared instruments can only
cover a fraction of the near-IR spectrum in a single observation,
and the broad FeII pseudo continuum can well cover the entire
observed spectral range, making it very difficult to recover
the true underlying continuum.
%The entire near-IR
%spectral region can be covered 
%with consecutive near-IR setups, but
%the inter-calibration between the multiple near-IR spectra is also very
%challenging, unless there is significant overlap between them,
%and even an uncertainty of 10\% in the intercalibration may
%result in an erroneous estimate of the continuum beneath the FeII
%bump.

In this letter the results of the analysis of near-IR spectra of 22 high
redshift quasars will be presented. These spectra were
obtained with an instrumental setup that can cover the full near-IR
spectrum from 0.8$\mu$m to 2.4$\mu$m in one shot (and with
very high throughput), therefore
allowing a good estimate of the continuum underlying the
FeII bump. Our observations provide a relatively large
sample of FeII/MgII measurements over a wide redshift range
and obtained with the same technique and fitting method. Therefore
these data allow for a reliable investigation of the redshift
dependence of the FeII/MgII ratio.
%at variance with previous studies
%which had to combine data obtained by different authors, with different
%instruments, and with different fitting techniques. 
Throughout the paper we adopt $\rm H_0=70~km~s^{-1}Mpc^{-1}$,
$\rm \Omega _M =0.3$ and $\rm \Omega _{\Lambda}=0.7$.

\vskip0.3cm
\figurenum{1}
%\begin{figure}[t]
%\plotone{f1.eps}
\centerline{\includegraphics[angle=0,width=8.5cm]{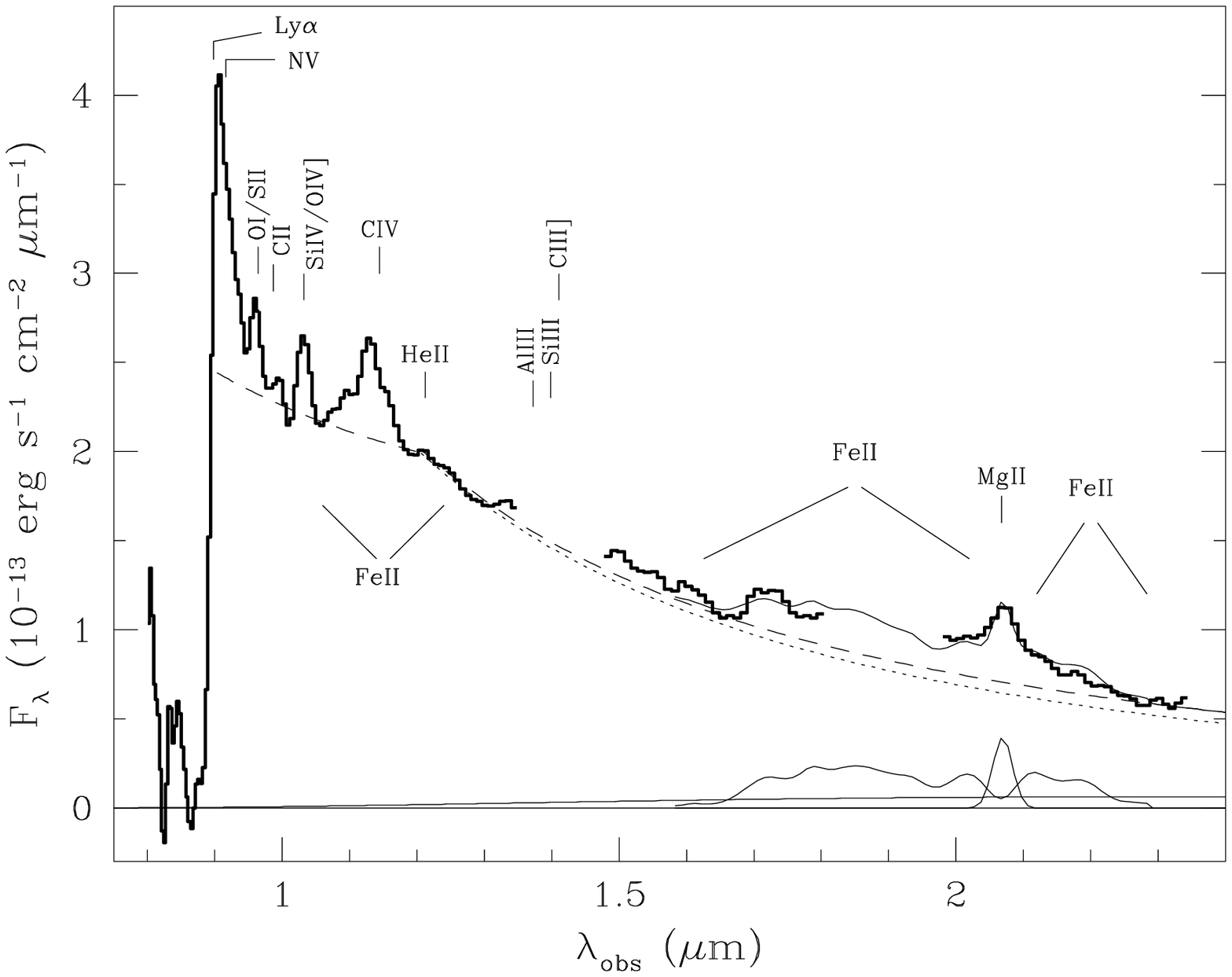}}
%\epsfig{file=qso64.eps}
%\figcaption{\footnotesize
\figcaption{
Near infrared spectrum of the highest redshift quasar
SDSSJ114816.64+525150.3 at z=6.4. The upper thick line shows the observed
spectrum, where regions of bad atmospheric transmission have
been omitted, while the overimposed thin solid line shows the best
fit to the continuum, FeII and MgII, as discussed in the text.
The dotted line shows the power-law component
($\rm F_{\lambda}\propto \lambda ^{\alpha}$, $\rm \alpha =-1.9$),
while the dashed line shows the total continuum fit (i.e. by
including the Balmer continuum).
The lower part of the figure shows the single components of the
fit and, more specifically, FeII bump, MgII line and Balmer
continuum.
\label{qso64}}
%\end{figure}

\section{Observations}

The observations were obtained with the
Near Infrared Camera Spectrograph (NICS) at the Italian Telescopio
Nazionale Galileo, a 3.56 m telescope. Among the various imaging and
spectroscopic observing modes \citep{bafet01}, NICS offers a
unique, high sensitivity, low resolution observing mode, which
uses an Amici prism as a dispersing element \citep{oliva03}. In this
mode it is possible to obtain the spectrum from 0.8$\mu$m to 2.4$\mu$m
in one shot. The throughput of the Amici prism is nearly two times higher
than other more commonly used dispersers.
The spectral resolution with a 0$''$.75 slit, as it
was in our case, is 75
(i.e. 4000~km~s$^{-1}$) and nearly constant over the whole
wavelength range. This observing mode is appropriate
to study the near-infrared continuum of faint sources,
and to detect broad ($\sim 5000$ km s$^{-1}$) emission lines
in faint quasars as well as pseudo-continua such as the FeII bump
which require a careful subtraction of the underlying continuum.
Observations were performed in three observing runs: 2002 November 7-9;
2003 February 25-29; and 2003 May 23-26. Typical integration
times ranged from $\sim$20 minutes, for bright objects, to
$\sim$2 hours for faint objects. We include also the data
of SDSSJ1044-0125 obtained previously at the same instrument
and already published
in \citet{maiet01}. Several quasars were observed more than
once on different nights to check for any instrumental or observational
artifacts in the individual spectra. Wavelength calibration was performed
by using an argon lamp and the deep telluric absorption
features. The telluric absorption was then removed by dividing
the quasar spectrum by a reference star spectrum (FV-GV) observed
at similar airmass. The intrinsic features and slope of the
reference star were then removed by multiplying the corrected quasar
spectrum by
a spectrum of the same stellar type from the library by \citet{pic98},
smoothed to our resolution. Absolute flux calibration was obtained
by using the photometry on the acquisition image or through photometry
reported in the literature.

In total 22 quasars were observed. These were selected among the
brightest (observable during each run) in each of the following
redshift ranges: 3.0-3.6, 4.3-5.1, 5.8-6.4, which ensure that the MgII
and a good fraction of the FeII are observed outside the deepest
atmospheric absorption regions (with the only exception of
SDSS1044-0125 at z$=$5.78 whose MgII is in a bad atmospheric band,
but which could be used to derive the average quasars spectrum, see
next section).

\vskip0.3cm
\figurenum{2}
%\begin{figure}
\centerline{\includegraphics[angle=0,width=8.5cm]{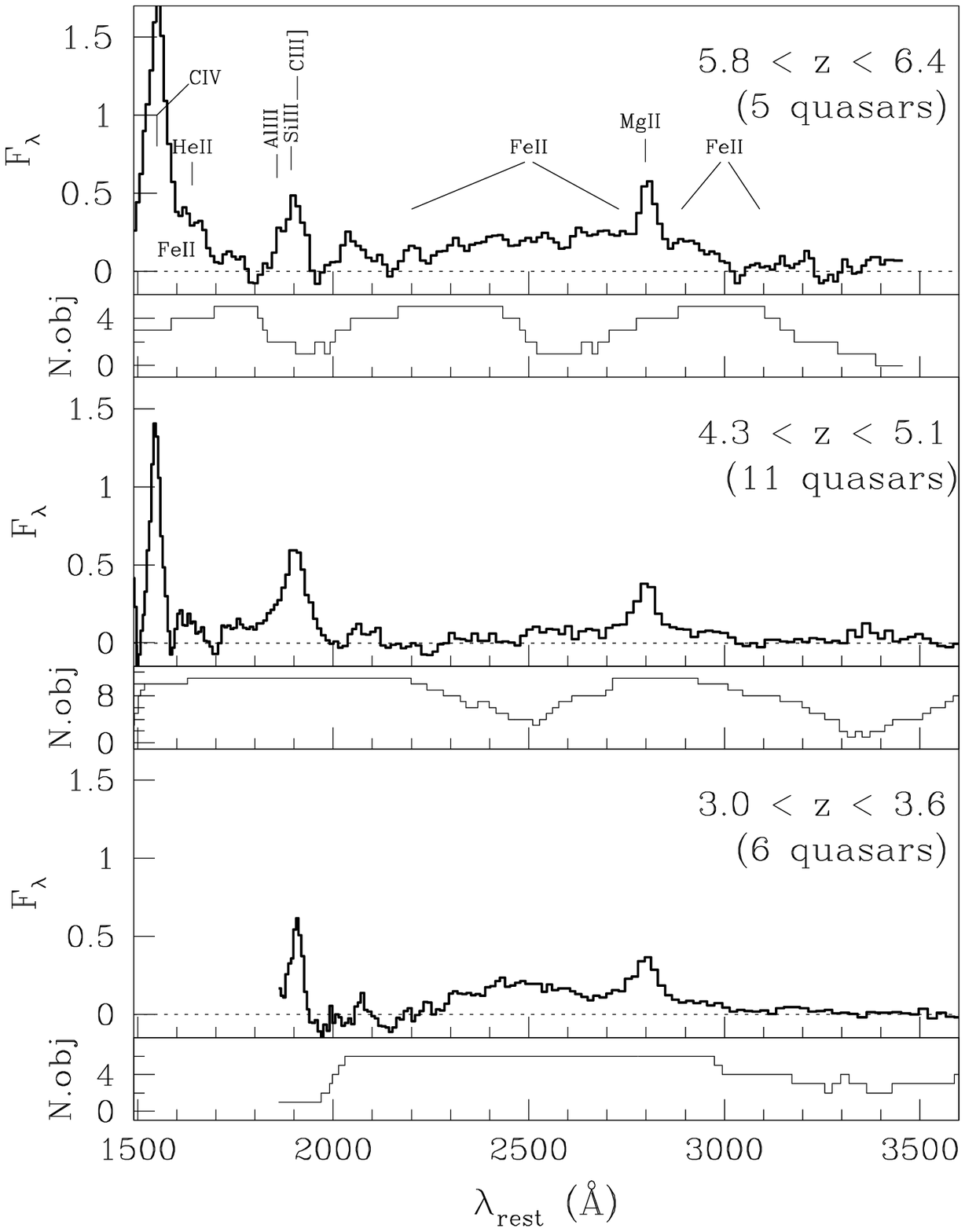}}
%\epsscale{.40}
%\plotone{f2.eps}
\figcaption{Average quasar spectra (residuals after continuum
subtraction) around the FeII UV bump obtained
by grouping the objects in the three main redshift bins and by
normalizing each spectrum to the continuum at the wavelength of
MgII. Beneath each panel we also show the number of individual
quasar spectra which where used to create the average spectrum.
Regions where the number of quasars is low correspond to the
spectral segments 
more seriously affected by the strong atmospheric absorption in that
redshift range (see text).
\label{avers}}
%\end{figure}

\section{Data analysis and fitting technique}\label{analys}

In Fig.\ref{qso64} we show the resulting spectrum of the highest
redshift quasar SDSSJ114816.64+525150.3, at z=6.40
\citep{fan03}\footnote{The best redshift fitting the observed wavelength
of MgII in our spectrum is 6.40, in agreement with the redshift
obtained by \citet{willot03} and \citet{barth03}.}, along with the
identification of the major spectral features.
Some blue-shift is observed for the
CIV line. This shift may be due to 
an imperfect correction of the atmospheric absorption
dip which occurs at the same wavelength; however, \citet{barth03} also
find the same effect, suggesting that the shift may be real.
The spectrum of the highest redshift quasar shows a prominent FeII bump.
All the other spectra and the detailed analysis of their
spectral features will be presented in a forthcoming paper. Here we focus
on the analysis of the UV iron complex and of the MgII~$\lambda$2798\AA\ blend.
The continuum fitting we performed was similar to the method adopted by 
\citet{dieet02,dieet03} and \citet{wills85},
i.e. we included both a powerlaw and a Balmer continuum.
The continuum was fitted
in the spectral regions free of emission features, and away from regions of
very bad atmospheric transmission.
In Fig.\ref{qso64} we show the power-law
($\rm F_{\lambda}\propto \lambda ^{\alpha}$, $\rm \alpha =-1.9$)
plus Balmer-continuum fit to the spectrum of SDSSJ114816.64+525150.3.
Note that in the highest redshift quasars (z$>$5.8) there is a flattening
in the continuum blueward of CIV, that is also observed in the optical
spectra of quasars at 3$<$z$<$5 \citep{francis91,vandenberk01},
possibly as a consequence of intervening metal absorption lines. At variance
with other authors, who prefer to fit the powerlaw by
sampling the continuum redward of 3100\AA\ and
the continuum between CIV and Ly$\alpha$ (but leaving positive continuum
residuals between CIV and FeII), we prefer to fit separately the
continuum on either side of CIV with two different power-laws. This also
ensures that the continuum fit of the quasars at z$>$5.8 is
consistent with the continuum fit of the lower redshift quasars.

In Fig.\ref{avers} we show the averages of the
quasars residual spectra after continuum subtraction (and normalized
to the continuum flux near the MgII line) and grouped
in the three main redshift ranges discussed above. In each
spectrum the regions of strong atmospheric absorption were
discarded, but the spread in redshift allows to cover continuously
the whole spectral region around FeII and MgII. Beneath each
average spectrum we show the number of quasar spectra which
populate each spectral region. While both the
low redshift and the high redshift average spectra clearly
show a prominent FeII bump, the average of the intermediate redshift
quasars is characterized by a significantly lower intensity of the FeII
bump (middle panel). This might be partly due to the region at
$\lambda < 2500$\AA \ being poorly populated since severe atmospheric
absorption separating the J and H bands affects several quasars in this
spectral region
at these redshifts. However, the FeII emission close to MgII is not
affected by this problem. Therefore we believe that the weakness of
FeII in the average spectrum of
intermediate redshift quasars is real.
%Moreover, the problem of a strong atmospheric
%absorption around the FeII bump is also present for the
%highest redshift average spectrum, which nonetheless show
%a prominent iron bump.

The MgII line and the residuals in the region around the FeII
bump (2200--3090\AA\ rest frame) were fitted, for each object,
by using the \citet[][]{vestwilk01} iron
template, smoothed to the velocity dispersion of MgII, and 
a broad gaussian at the wavelength of the MgII line.
These fitting components
for the case of
SDSSJ114816.64+525150.3 are shown in Fig.\ref{qso64}. The FeII flux
was integrated between the rest-frame wavelengths 2200 and 3090\AA .
In Table~\ref{tab1}
we report the FeII/MgII ratio for all quasars observed by us. We also list
the rest-frame luminosity $\lambda$L$_{\lambda}$ at 1450\AA\ (note that thanks
to our wide spectral coverage we can measure this quantity directly in several
quasars, or by performing only a minor extrapolation).

We note that 
the continuum and Fe fitting techniques, as well as the instrumental
setup, have often been different among previous works on this subject.
As a consequence, a comparison of the results between these different
studies may be subject to strong uncertainties
\citep[although some authors attempt to correct as much as possible
for these systemic differences, e.g.][]{dieet03}.
Within this context we emphasize that
our survey provides uniform and homogeneous
FeII/MgII 
measurements for a set of 22 quasars
spanning a wide redshift range (3.0--6.4) obtained with the
same instrumental setup (which covers the entire near-IR in one
spectrum), same spectral
resolution (constant over the whole spectral
range) and same fitting technique. Therefore they represent a
self-consistent sample which is well suited to
study the trend of FeII/MgII with redshift.

Some of the quasars observed by us were also observed by other authors
and deserve some comparison, even with the caveats
discussed above. BR~2237-0607 and BR~0019-1522 were observed
by \citet{dieet03} who find higher FeII/MgII ratios than us, but
the inconsistency is not large when errors are taken into account.
BR~2237-0607 was also observed by \citet{iwaet02}, who obtain a ratio
consistent with ours.
SDSSJ103027.10+052455.0 was observed with HST/NICMOS
by \citet{freet03} and they derived a much lower value for the 
FeII/MgII ratio ($2.1\pm1.1$) than found by us ($8.64\pm 2.47$).
However, Freudling
et al. did not measure MgII directly in this object,
but instead estimated the MgII flux from the measured CIII] flux and an
assumed CIII]/MgII ratio similar to the
average of other quasars. Also, they did not sample the
continuum redward of the FeII bump.
SDSSJ114816.64+525150.3 was observed by \citet{barth03},
who found a lower FeII/MgII value ($4.7\pm 0.4$) than ours, though
marginally consistent with errors ($\sim 1.5\sigma$). Their spectral
coverage does not extend enough to sample the continuum outside the
iron bump, this might have led them to overestimate the underlying continuum.

%The FeII/MgII ratio as a function of redshift is shown in Fig.\ref{femg}
%for the objects observed by us.
In Fig.\ref{femgbin} we show the
average of the FeII/MgII ratios within the three redshift
ranges discussed above.
In the same figure we also plot
the FeII/MgII ratio obtained by \citet{dieet03} from composite optical spectra of quasars
at lower redshift matching nearly the same
luminosity range as the objects in our sample.
No trend of the FeII/MgII with
luminosity was found within the luminosity range of our sample,
which spans about 1.5 orders of magnitude.

\vskip0.3cm
\figurenum{3}
%\begin{figure}
\centerline{\includegraphics[angle=0,width=7.5cm]{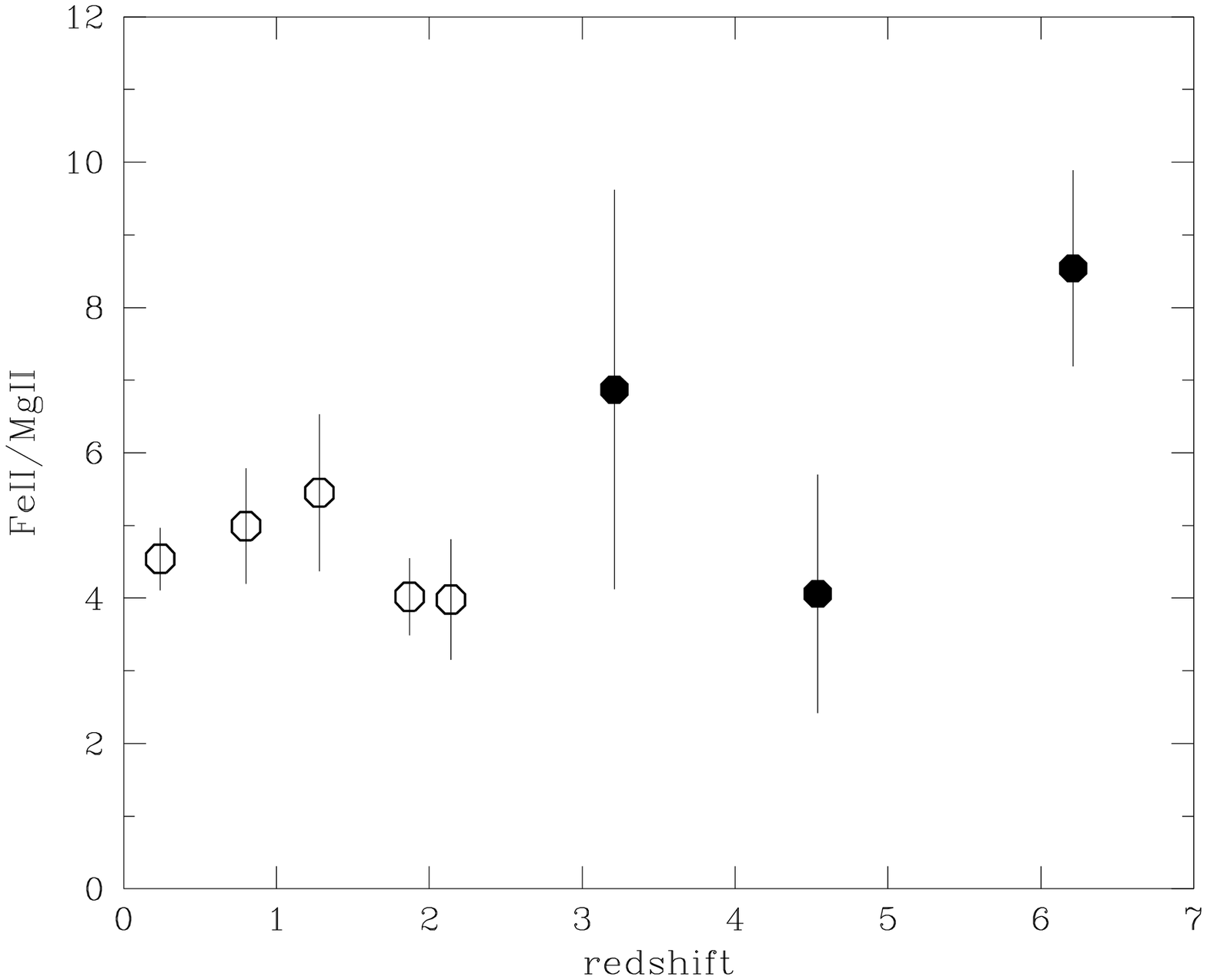}}
%\plotone{f3.eps}
\figcaption{Filled symbols show the average of
FeII/MgII ratios of the quasars in our three main redshift bins.
Hollow symbols show the FeII/MgII ratio
obtained by \citet{dieet03} from optical composite spectra of quasars
at lower redshift.
\label{femgbin}}
%\end{figure}

\section{Discussion}\label{disc}

Some evidence for iron super-solar abundance was found by \citet[][]{wills85}
in low to intermediate redshift quasars. In our spectra we detect
iron emission also in all high redshift quasars.
The ratio of FeII/MgII is nearly constant over the redshift range zero
to 6.4.
The analysis of the 
high redshift quasars indicate a possible increase in the FeII/MgII ratio at 
redshifts $\rm z\sim 6$.
If the FeII/MgII flux ratio is tracing
the Fe/$\alpha$ abundance ratio, this result would imply that the iron abundance
is nearly constant or even slightly increasing from
redshift zero to 6.4, i.e. when the age of the universe was about 800~Myr.
While this finding is difficult to reconcile with
the scenarios expecting a delay of 1~Gyr or more for the iron enrichment, models
of rapid and strong star formation at high redshift offer a plausible explanation.
Indeed, \citet{matrec01} have shown that
models of early formation of ellipticals, with very efficient star formation
and lasting $\sim$0.4~Gyr, imply a peak of SNIa production
(hence of iron enrichment)
already at 0.3~Gyr after the onset of star formation. They have also shown
that in the case of an instantaneous burst
the SNIa production peaks at only 50~Myr after the burst.
As a consequence, our results can be explained with two possible scenarios:
\begin{enumerate}
\item A major episode of star formation in these highest-z quasar hosts occurred
    at z $\geq$ 9, to allow the estimated 0.3~Gyr required for
    SNIa to enrich the ISM within the context of models for star 
    formation in elliptical hosts. The assumption that high redshift
    quasars are hosted in young elliptical galaxies may be reasonable,
    based on the studies of quasar hosts at lower redshift \citep{kuku01,dunlop03}.
    Nonetheless, it is also important to keep in mind that there
    is no direct observational evidence that
    these highest-z quasars reside in "normally" evolving 
    elliptical galaxies.
\item An ``instantaneous'' burst of star formation occurred as late as z $\sim$ 7,
    and SNIa had enough time to enrich the ISM within $\sim$ 50~Myr of the burst.
    The ideal case of a
    single instantaneous burst is not likely to represent the scenario
    of spheroids and galaxies formation in general, for which a star formation extended in
    time is more plausible. However, a single instantaneous burst may have
    occurred in the nuclear region of these quasars, causing a rapid iron
    enrichment limited mostly to the central region.
\end{enumerate}

It is interesting to note
that the first scenario (early formation of ellipticals) implies a redshift for the
first major episode of star formation (z$\geq$9) which 
is consistent with the redshift of the re-ionization (at z$\approx 20\pm 10$)
inferred from the recent Wilkinson Microwave Anisotropy Probe (WMAP)
results \citep{benet03}. While the second
scenario would be more consistent with the secondary re-ionization (at
z$\sim$6) proposed by \citet[][]{cen03} and \citet[][]{wylo03}.

We also find a population of quasars at z$\sim$4.3--5.1 with low iron emission
(Table~\ref{tab1}). These objects are
responsible for the apparently lower FeII bump in the average spectrum
of the quasars in this redshift bin (Fig.\ref{avers}). Other quasars with
such low Fe emission in this redshift range were also found by \citet{dieet03}
and by \citet{iwaet02} (see also \citealt{fiore03}).
Statistically we cannot make any strong statement
on the fraction of quasars with low FeII. However, they could represent a population
of quasars hosted in stellar systems which evolved more slowly than the quasars
with higher Fe emission.

An alternative explanation for the FeII/MgII trends discussed above is that
the physical conditions of the Broad Line Region in high redshift quasars
are significantly different than in quasars at lower redshift. In this case
the observed FeII/MgII ratio at high redshift
would reflect the combined effects of abundance variations {\it and}
variations of other physical parameters such as
density, ionization parameter, and microturbulence.
Disentangling these effects requires our observations to be complemented
with the detection of the optical iron bump \citep{veret03},
which is shifted beyond the
K-band at z$>$5. Observations sensitive enough to detect the optical
iron bump at these redshifts will probably have to await the James Webb
Space Telescope (JWST).

\acknowledgments

We thank the referee M. Dietrich for helpful comments.
Part of this work was developed during the Guillermo Haro Workshop in 2003.
RM acknowledges partial financial support from the Italian Space Agency (ASI).
YJ and RM acknowledge support from CONACyT grant J32178-E.

\clearpage

\begin{deluxetable}{lccc}
%\tabletypesize{\scriptsize}
\tablecaption{FeII/MgII measurements \label{tab1}}
\tablewidth{0pt}
\tablehead{
\colhead{Name} & \colhead{z}   & \colhead{FeII/MgII}   &
\colhead{log $\lambda$L$_{\lambda}$(1450\AA)}
}
\startdata
 BR 0019-1522     & 4.52    &   2.02$\pm$1.01   &   46.76\\
 PSS J0134+3307   & 4.53    &   1.68$\pm$0.42   &   46.87\\
 SDSS J021102.72-000910.3 & 4.73    &   4.10$\pm$2.05   &   46.17\\
 SDSS J075618.14+410408.6   & 5.09    &   3.43$\pm$1.37   &   46.54\\
 SDSS J103027.10+052455.0   & 6.28    &   8.65$\pm$2.47   &   46.68\\
 SDSS J104433.04-012502.2   & 5.78    &   ---\tablenotemark{a}   &   46.88\\
 SDSS J104845.05+463718.3   & 6.23    &   8.03$\pm$1.22   &   46.81\\
 SDSS J114816.64+525150.3   & 6.40    & 8.20$\pm$2.10   &  46.98\\
 SDSS J130608.26+035626.3   & 5.99    &   9.03$\pm$2.26   &   47.32\\
 GB 6 15007+4848    & 3.20    &   4.17$\pm$0.63   &   47.63\\
 BR J1603+0721     & 4.38    &   4.06$\pm$0.81   &   46.86\\
 SDSS J160501.21-011220.6   & 4.92    &   5.30$\pm$1.35   &   46.46\\
 PSS J1633+1411    & 4.35    &   3.23$\pm$1.28   &   47.02\\
 HS 1649+3905      & 3.05    &   7.40$\pm$0.74   &   46.72\\
 SDSS J173352.23+540030.5   & 3.42    &   7.74$\pm$1.16   &   46.99\\
 SDSS J173744.88+582829.6   & 4.94    &   3.55$\pm$1.42   &   46.75\\
 S5 1759+75        & 3.05    &   5.01$\pm$2.01   &   47.41\\
 PSS J2154+0335    & 4.36    &   5.83$\pm$1.75   &   47.81\\
 SDSS J220008.66+001744.8    & 4.77    &   5.50$\pm$1.65   &   46.56\\
 BR 2237-0607      & 4.56    &   3.00$\pm$1.20   &   47.42\\
 LBQS 2231-0015     & 3.02    &   5.77$\pm$0.86   &   47.06\\
 SDSS J234625.67-001600.5   & 3.50    &   11.7$\pm$2.92   &   46.90\\
\enddata
\tablenotetext{a}{The MgII of this object was not measured because it
is located
in the atmospheric absorption between H and K. However this quasar was used
to obtain the average spectrum in Figure~\ref{avers}.}
\end{deluxetable}

\end{document}